# Black-phosphorus Terahertz photodetectors

*Leonardo Viti[1], Jin Hu[2], Dominique Coquillat,[3] Wojciech Knap,[3,4] Alessandro Tredicucci,[5] Antonio Politano[6] and Miriam S. Vitiello\*[1]*

[1] NEST, Istituto Nanoscienze – CNR and Scuola Normale Superiore, Piazza San Silvestro 12, Pisa, I-56127

[2] Department of Physics and Engineering Physics, Tulane University, New Orleans, LA-70118, USA

[3] Laboratoire Charles Coulomb (L2C), UMR 5221 CNRS-Université de Montpellier, Montpellier, F-France

[4] Institute of High Pressure Institute Physics Polish Academy of Sciences Warsaw Poland

[5] Dipartimento di Fisica, Università di Pisa, Largo Pontecorvo 3, 56127 Pisa, Italy

[6] Università degli Studi della Calabria, Dipartimento di Fisica, via Ponte Bucci, 87036 Rende (CS), Italy

\* miriam.vitiello@sns.it

Keywords: inorganic two-dimensional (2D) materials, nano-transistors, photodetectors

**Abstract.** The discovery of graphene and the related fascinating capabilities have triggered an unprecedented interest in inorganic two-dimensional (2D) materials. Despite the impressive impact in a variety of photonic applications, the absence of energy gap has hampered its broader applicability in many optoelectronic devices. The recent advance of novel 2D materials, such as transition-metal dichalcogenides or atomically thin elemental materials, (e.g. silicene, germanene and phosphorene) promises a revolutionary step-change. Here we devise the first room-temperature Terahertz (THz) frequency detector exploiting few-layer phosphorene, e.g., a 10 nm thick flake of exfoliated crystalline black phosphorus (BP), as active channel of a field-effect transistor (FET). By exploiting the direct band gap of BP to fully switch between insulating and conducting states and by engineering proper antennas for efficient light harvesting, we reach detection performance comparable with commercial detection technologies, providing the first technological demonstration of a phosphorus-based active THz device.

Graphene-oriented [1] research has had a dramatic impact during the last decade.[2] The superior carrier mobility, induced by the massless Dirac fermions in graphene, combines with a gapless spectrum that, although beneficial for applications requiring frequency-independent absorption, [3,4] also prevents the effective switching of its conductivity in electronic devices and the achievement of a high on/off current ratio in transistors. Finite and direct bandgaps are desirable for a wealth of applications, including transparent optoelectronics, photovoltaics and photodetection. As an example, current visible graphene detector performances are strongly limited by the large dark currents that dominate under non-zero bias operation. [5,6]

These issues are driving present research in the quest for alternative 2D materials: as a prototypical example, single-unit-cell thick layers of transition-metal dichalcogenides (TMDCs: $MoS_2$, $MoSe_2$, $WS_2$, $WSe_2$, etc.) have recently emerged as a valuable alternative. [7] 2D TMDCs can be obtained from bulk crystals by employing the micromechanical exfoliation method, [4] like for the case of graphene, but they show a direct bandgap, ranging between ~0.4 eV and ~2.3 eV, which enables applications that well complement graphene capabilities.[8] In particular, 2D TMDCs are suitable for photovoltaic applications [9] and for devising robust ultra-thin-body field effect transistor (FET) architectures which can easily provide subthreshold swing of ~ 60 mV/dec and $I_{on}/I_{off}$ ratio up to $10^8$. [10] Nonetheless, their relatively low mobility ($\leq 200$ $cm^2V^{-1}s^{-1}$) is a major constraint for high-frequency electronic applications.

A good trade-off between graphene and TMDCs is represented by a novel class of atomically thin 2D elemental materials: silicene, [11] germanene [12] and phosphorene. [13] Among them, the latter one allows a peculiar single- or few-layer isolation from its bulk phase, e.g., black-phosphorus (BP). Unlike silicon and germanium, BP, the most thermodynamically stable allotrope of the phosphorus element in standard conditions, shows a layered graphite-like structure, where atomic planes are held together by weak Van der Waals forces of attraction, thus allowing the application of standard micromechanical exfoliation techniques.

As in graphene, each atom in BP is connected to three neighbours, forming a stable layered honeycomb structure with an interlayer spacing of ~5.3 Å. In contrast with graphene, the hexagonally distributed phosphorus atoms are arranged in a puckered structure rather than in a planar one (**Figure 1a**). This property generates an intrinsic in-plane anisotropy that results in a peculiar angle-dependent conductivity. [14]

Bulk BP has a small direct bandgap of ~0.3 eV, which enables the complete switching between insulating and conducting states in transistor devices. The reduction of the flake thickness leads to quantum confinement which enhances the gap up to $E_g$~1.0 eV in the limit case of phosphorene (a single layer of BP); as a result the $I_{on}/I_{off}$ ratio of a BP-based FET can be improved by employing thinner flakes: an $I_{on}/I_{off}$ ratio of ~$10^5$ has been recently reported [8] in back-gated FET structures. On the other hand, thickness reduction is detrimental for carrier mobility: thinner flakes are more vulnerable to scattering by interface impurities [15] and the effective mass of charge carriers increases when the number of atomic layers is reduced. [16] Despite this, BP thin films are endowed with hole mobilities exceeding 650 cm$^2$V$^{-1}$s$^{-1}$ at room temperature (RT) and well above 1000 cm$^2$V$^{-1}$s$^{-1}$ at 120 K, [14] thus overtaking the limiting factor of large-gap TMDCs, allowing to reach high-frequency operation up to 20 GHz. [18] For all the reasons above, BP represents an ideal material for infrared optoelectronic applications [17] and high-speed thin film electronics. [18]

Furthermore, the superb $10^5$ $I_{on}/I_{off}$ ratio of BP-based FETs makes BP well suited for detection of Terahertz (THz) -frequency light, being finite and direct bandgaps and huge carrier density tunability undisputed benefits in the viewpoint of higher modulation frequency, large responsivity ($R_v$) and low dark currents.

THz detection in FETs is based on the rectification of plasma waves induced by the external *ac* electric field. [19] When an electromagnetic field is coupled between the source (S) and the gate (G) electrodes, it excites carrier density oscillations, which, in turn, generate a driving longitudinal electric field through the channel. This simultaneous modulation of carrier density and drift velocity results in the onset of a *dc* signal that can be measured at the drain (D) electrode as a voltage, if D is kept in an open circuit configuration (photovoltage mode), or as a current, if D is in a short-circuit configuration (photocurrent mode). The asymmetric feeding of the ac signal, [20] or the intrinsic asymmetry in the transistor channel, [21] defines a preferential direction for charge flow across the channel, allowing FET detectors to work without any applied bias, thus reducing the noise level and the system complexity.

At RT the propagation of plasma waves in the transistor channel is usually limited by the phonon scattering which reduces the channel mobility ($\mu$): when the plasma oscillations, excited at the S side of the transistor, are damped in a distance $L_d$, which is shorter than the channel length ($L_c$), the FET is said to be operating in the non-resonant overdamped regime. This mechanism is very promising for applications because the rectification takes place over a distance $L_d$ regardless of the total length of the channel: the FET can operate beyond its cut-off frequency even though the channel mobility is low and its dimensions are not sub-micron. FETs allow for the integration, as active channels, of a variety of one-dimensional [22,23] or 2D [11,13,24-26] structures whose properties can either be exploited for [22,25] or investigated via [27] THz detection experiments.

Conceiving and exploiting new material combinations can open the path to ground-breaking implementations of active devices and passive components across the intriguing and underexploited THz frequency range. Here we show efficient THz detectors working at RT by combining top-gated BP FETs with THz antennas designed to enhance sensitivity.

BP crystals were grown via chemical vapour transport techniques. [28] Flakes having thickness ≤10 nm were then mechanically exfoliated from bulk BP crystal using a standard adhesive tape technique on a 300 nm thick $SiO_2$ layer on the top of a 300 μm intrinsic silicon wafer. The thickness of individual BP flakes was assessed via a combination of optical microscope mapping (**Figure 1b**), scanning electron microscopy imaging (**Figure 1c**) and atomic force microscopy (AFM) imaging (Fig. 1e). Reproducible correlation has been found between the color map of the flake, as seen under the optical microscope (Fig. 1b), and its thickness measured via atomic force microscopy (AFM) (**Figure 1d**).

Micro-Raman spectroscopy measurements have been performed to evaluate the crystalline quality of BP flakes: the Raman spectrum presents three characteristic peaks at 362, 440, and 468 cm$^{-1}$ (**Figure 1d**) corresponding to the $A_g^1$, $B_{2g}$, and $A_g^2$ phonon modes observed in bulk BP. The topographic AFM scan of a single transferred flake (**Figure 1d**) shows a layered structure with two visible stacks with thicknesses 6.2 and 1.7 nm, respectively. By measuring a set of similar flakes we found that every stack has a thickness corresponding to an integer multiple of $\sim 0.61$ nm, i.e. the thickness of a BP monolayer (phosphorene).

To devise the FET detectors, we selected 10 nm thick BP flakes. This choice is motivated by the fact that THz detection, in the overdamped plasma-wave regime, is only marginally affected by the channel mobility, that decreases by reducing the flake thickness; conversely, the expected photoresponse significantly increases when huge carrier tunabilities are reached, i.e. in thinner flakes. THz detectors have been realized by exploiting a combination of electron beam lithography (EBL) and metal evaporation (see Experimental details). The S and G electrodes were patterned in the shape of two halves of a planar bow-tie antenna having a total length 2L = 500 µm and a flare angle of 90°, in resonance with a 0.3 THz radiation. **Figure 2** shows the device layout: the channel length (source-to-drain distance) has been set to $L_c$ = 2.78 µm, the gate length is $L_G$ = 580 nm and the average channel width W = 2.6 µm. Under this configuration and in the presence of an 80 nm thick oxide layer, we simulated the geometrical gate-to-channel capacitance ($C_{gc}$) with a commercial 3D-FEM simulation software (COMSOL Multiphysics) and found $C_{gc} \sim 1.1$ fF.

The RT transport characterization was carried out using two *dc* voltage generators to drive the source-to-drain voltage ($V_{SD}$) and the gate voltage ($V_G$) independently (**Figure 3**). The device shows ohmic behaviour, with no signature of a contact Schottky barrier at RT .The transconductance characteristics have been acquired while sweeping $V_G$ in a limited bias range (see **Figure 3**) to prevent the top-gate insulating layer from breakdown. The resulting device is a p-type depletion mode FET with a visible hysteresis connected with the direction of the gate sweep. This effect is likely due to charge trapping in the top oxide layer and at the BP-dielectric interfaces. [29] The ambipolar transport, typical of BP-based FETs, [15,30] has not been observed in the present case due to the use of Cr/Au contacts that can significantly alter the band alignment and eventually suppress the electron conduction even at high positive $V_G$. [31] The achieved maximum transconductance $g_m = 150$ nA/V is independent of the gate sweep direction (dash-dot line in **Figure 3a**) and the $I_{on}/I_{off}$ ratio ~$10^3$, provides a satisfactory switching behavior.

In a 2D FET the mobility is conventionally extrapolated in a back-gate (BG) configuration [10,30] or via capacitance-to-$V_G$ (C-$V_G$) measurements. [29] In the present configuration, these strategies are difficult to be implemented because: (i) back-gates require doped substrates, which are detrimental for THz detection; (ii) any C-$V_G$ measurement is largely affected by the huge shunt capacitance induced by the presence of the bow-tie antenna. Therefore, here we have adopted a different method to extrapolate the mobility ($\mu_{FE}$) indirectly from the transconductance $g_m$ curve through the relation $g_m = \mu_{FE} C_{ox} V_{SD} W / L_G$, where $C_{ox} = C_{TG}/A_G$ is the oxide capacitance per unit area, $C_{TG}$ is the top-gate capacitance and $A_G$ the gated area. This relation holds while the mobility is constant over the entire length and does not depend on the applied bias, as is reasonable to assume in a top-gated system. For a thin layer channel, $C_{TG}$ (**Figure 3b**, inset) is given by the sum in series of the geometrical capacitance $C_{gc}$ (whose value can be obtained via FEM simulations) and the parallel of the interface trap capacitance ($C_t$) and the quantum capacitance ($C_q$): [29,32]

$$C_{TG} = \left( \frac{1}{C_{gc}} + \frac{1}{C_t + C_q} \right)^{-1} \qquad (1)$$

The quantum capacitance is related to the material density of states and takes into account the fraction of $V_G$ dropped within the channel to modify the carrier population. Therefore, according to equation (1), the gate voltage modulates the band structure (via $C_{gc}$) and simultaneously fills trap states (via $C_t$) and carrier states (via $C_q$). An estimation of the quantity ($C_t + C_q$) can be obtained from the subthreshold swing ($S_s$) of the FET. The logarithmic plot of the transconductance curve shows a linear region (shaded areas in Fig.3b) whose slope corresponds to the subthreshold slope $(S_s)^{-1}$: this regime corresponds to the onset of a pure hole thermionic emission current over the potential barrier generated at positive $V_G$. In general, $S_s \sim 60\ \beta$ mV/dec, where $\beta$ is the band movement factor: $\beta = [1+(C_t+C_q)/C_{gc}]$. In the subthreshold regime, the channel is almost depleted of free carriers, hence $C_q$ can be safely neglected: the channel bands move one-to-one with the applied gate bias. [8] In the present case, we estimate $\beta = 7$ (independently from the $V_G$ sweep direction) from the linear fit to the data (shaded areas, **Figure 3b**), then $C_t = 6\ C_{gc}$; by varying $V_G$, the top gate capacitance will vary in the range [6/7 $C_{gc}$, $C_{gc}$]. From these considerations, we can infer a lower bound for $\mu_{FE}$:

$$\Delta u_T \propto -\frac{1}{\sigma} \cdot \frac{d\sigma}{dV_G} \cdot \left[\frac{R_L}{\frac{1}{\sigma}+R_L}\right] \tag{2}$$

corresponding to $\mu_{FE} = 470$ cm$^2$V$^{-1}$s$^{-1}$.

In order to prove THz detection at RT, we employed an electronic source whose output frequency was tuned in the range 0.26-0.38 THz. The output beam, having polarization parallel to the antenna axis, was collimated by a set of two 1/f off-axis parabolic mirrors in a 4 mm diameter spot while its amplitude was modulated as a square wave at 618 Hz by means of a mechanical chopper. The photoresponse was measured in photovoltage mode: with the S electrode grounded, the response voltage ($\Delta u$) was collected at the D electrode in open circuit configuration (see Methods).

**Figure 4a** shows the channel conductance dependence on the gate bias, measured while the 0.298 GHz radiation (see Experimental details) is impinging on the device. When the detection mechanism is dominated by the aforementioned plasma-wave rectification, the transfer characteristics allow predicting the device responsivity, according to the diffusive hydrodynamic model of transport. [6,19] Loading effects arising from the finite impedance ($R_L$) of the measurement setup have been also properly weighted, leading to:

$$\Delta u_T \propto -\frac{1}{\sigma} \cdot \frac{d\sigma}{dV_G} \cdot \left[ \frac{R_L}{\frac{1}{\sigma} + R_L} \right] \qquad (3)$$

where the minus sign accounts for the hole majority carriers. Figure 4b shows the predicted $\Delta u_T$ trend as a function of $V_G$: the THz response is expected to peak around $V_G = +3.5$ V. The comparison with the experimental responsivity curve (**Figure 4c**) reveals a good agreement and unveils that THz-induced plasma waves arise in the BP channel, allowing the efficient detection of the rectified signal. In the present experiment, plasma-waves are strongly damped being $2\pi\nu\tau \ll 1$. Here $\tau$ is the hole relaxation time whose upper limit of $\approx 200$ fs has been assessed through the mobility relation $\mu = e\tau/m^*$, where $m^*$ is the hole effective mass. For BP flakes thicker than 5 layers, $m^*$ is smaller than the electron mass, regardless of the flake orientation. [16]

It is worth noticing that, at negative gate biases, $R_v$ does not decrease to zero (Fig. 4c), in contrast with the model predictions, but saturates at an average value of $\sim 0.03$ V/W, likely due to the enhanced impedance matching between the antenna (70 $\Omega$) and the FET. The measured net residual noise, shown in **Figure 4c** (black curve), has been estimated by measuring the photovoltage signal while obscuring the THz beam with an absorber. The direct comparison between the signal curve (blue) and noise curve (black), the maximum signal-to-noise ratio (SNR) was estimated to be ~32 at $V_G = 3.26$ V.

The noise-equivalent power (NEP) has been extracted from the ratio $N_{th}/R_v$, by assuming that the main contribution to the noise figure is the thermal Johnson-Nyquist noise ($N_{th}$), [25] associated with the non-zero resistance of the FET channel. [33] Although this hypothesis neglects the 1/f and the shot noise contributions, it provides a lower limit for the NEP. [34,35] Minimum values of ~ 40 nW/(Hz)$^{1/2}$ have been reached, comparable with the NEP of graphene-based THz photodetectors. [25] The achieved results highlights the potential of the devised BP active THz devices to deeply impact a plethora of technological applications in the gap between microwave and optical technology domains.

**Experimental Details**

*Fabrication* Micro-Raman spectroscopy on the exfoliated BP flakes have been performed with a Renishaw (InVia) system, equipped with a frequency doubled Nd:Yag 532 nm laser having maximum output power of 500 mW (CW). In the present experiments, we kept the optical intensity ≤ 0.4 mW/μm$^2$, since intensity values larger than 0.8 mW/μm$^2$ are likely to damage the flakes. The S-D FET channel has been defined via a combination of electron beam lithography (EBL) and thermal evaporation to deposit the (10/70 nm) (Cr/Au) S and D metal contacts. A 80 nm think SiO$_2$ oxide layer was then deposited on the sample via Ar sputtering. The G electrode was patterned on the oxide via EBL and a (10/90 nm) (Cr/Au) layer thermally evaporated on it. Although high work-function metals, like Pd or Ni, are preferable to achieve ohmic contacts on p-type semiconductors [30], chromium ensures better adhesion to the SiO$_2$ substrate, which turned out to be crucial to safely lift-off large area planar metallic structures. Furthermore, the electrical performances of BP-based FETs are prone to time degradation due to ambient air humidity, [36] therefore surface passivation strategies have to be adopted. The deposition of an insulating oxide layer over the exposed face of a BP-flake proved to be reasonably effective, [37] in our case. We indeed electrically characterized our FETs after three weeks of exposure to the environment, finding a less than a factor of 4 reduction in the channel conductance and an I$_{on}$/I$_{off}$ ratio two times smaller than in post processing tests. Further improvements can be envisaged if SiO$_2$ is replaced by high-k dielectrics (Al$_2$O$_3$, HfO$_2$, etc.), which are expected to provide more efficient protections.[37]

*Optical testing* The voltage response ($\Delta u$) has been recorded at the D electrode in an open circuit configuration, while keeping the S electrode grounded, by means of a lock-in amplifier, connected to a low-noise voltage pre-amplifier having an input impedance of 10 MΩ and a gain factor $G_n$=1000. $\Delta u$ has been retrieved from the voltage signal read on the lock-in (LIA) via the relation $\Delta u$ = 2.2·LIA/$G_n$, where the factor 2.2 accounts for the square wave modulation, since the lock-in measures the rms of the sine wave Fourier component. The highest photoresponse was achieved at 0.28 THz; the corresponding source power, measured at the focal point of the optical system was 325 μW. The detector responsivity ($R_v$) was then extracted from $\Delta u$ through the relation $R_v$ = ($\Delta u \cdot S_t$)/($P_{THz} \cdot S_a$), where $S_t$ is the beam spot area ($S_t = \pi d^2/4$, where d is the spot diameter) and $S_a$ is the active detection area. In our geometry, the area of the 500 μm long bow-tie antenna is lower than the diffraction limited one ($S_\lambda$), hence we assumed $S_a = S_\lambda = \lambda^2/4$.

## Acknowledgments

This work has been partially supported by the Italian Ministry of Education, University, and Research (MIUR) through the program FIRB - Futuro in Ricerca 2010 RBFR10LULP "Fundamental research on Terahertz photonic devices" and by the European Union through the MPNS COST Action"MP1204 TERA-MIR Radiation: Materials, Generation, Detection and Applications". W.K acknowledges the National Science Poland Centre (DEC-2013/10/M/ST3/00705). JH is supported by NSF/LA-SiGMA program under award #EPS-1003897. AP thanks Anna Cupolillo for helpful discussions.

**Figure Captions**

**Figure 1: Flake identification and characterization. (a)** BP atoms are arranged in puckered honeycomb layers bounded together by Van der Waals forces. **(b)** Optical image of exfoliated flakes of BP. **(c)** Scanning electron microscope (SEM) image of the same area; optical and electron-beam microscopy were used as a first step to identify thinner layers of BP. **(d**) Micro-Raman spectrum measured using a 523 nm excitation laser: peaks are found at 362, 440, and 468 $cm^{-1}$, corresponding to the $A_g^1$, $B_{2g}$, and $A_g^2$ vibrational modes, respectively. **(e)** Atomic force microscopy topographic image of an individual flake with thickness 6.2 nm. A topographic line profile, acquired along the dashed green line is shown.

**Figure 2: Sample Fabrication**. **(a)** Sketch of the device structure (vertical section). **(b)** False colors SEM image of the BP-based FET. The channel length ($L_c$) is 2.7 µm, the gate length ($L_g$) is 530 nm. **(c)** S and G electrodes are designed to form a 500 µm, 90° flare angle, planar bow-tie antenna. The D electrode is connected to a rectangular bonding pad.

**Figure 3. Transport characterization. (a)** RT transfer characteristic obtained while sweeping $V_G$ in the (-3 V; 4 V) range, by keeping $V_{SD}$ = 1 mV. $I_{SD}$ was amplified by a factor $10^6$ by using a transimpedance amplifier. The shaded areas indicate the linear regime of operation. From a linear fit to the data (dash-dot-lines) we obtained $g_m$ = 150 nA/V. **(b)** Logarithmic plot of $I_{SD}$ as a function of $V_G$. The shaded areas mark the sub-threshold regime and the dashed lines are linear fits to the data, used to infer the sub-threshold slope.

**Figure 4. THz detection. (a)** Channel conductance (σ) as a function of $V_G$ measured while progressively increasing $V_G$ and with the 0.298 THz radiation impinging on the detector surface. **(b)** Predicted photoresponse as a function of $V_G$, under the overdamped plasma-wave regime. This is the expected responsivity following a diffusive model of transport. **(c)** Gate bias dependence of the experimental RT responsivity. The red line (curve 1) was measured by impinging the THz beam on the detector surface; the black line (curve 2) was measured while blanking the beam with an absorber (considering unaltered the incident power). The extracted maximum signal to noise ratio (SNR) is ~20 for $V_G$ = -2 V. Inset: NEP as a function of $V_G$, extracted from the relation $N_{th}/R_v$ . The spectral density of $N_{th}$ has been calculated via the relation $N_{th} = (4k_BT\sigma^{-1})^{\frac{1}{2}}$.

Figure 1

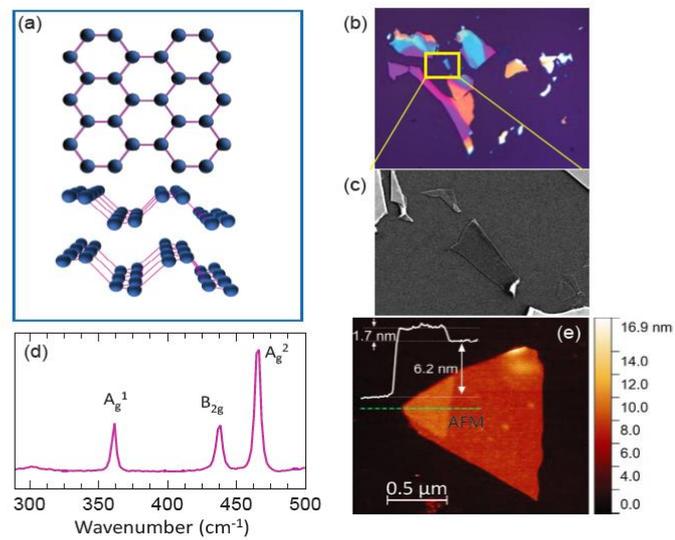

Figure 2

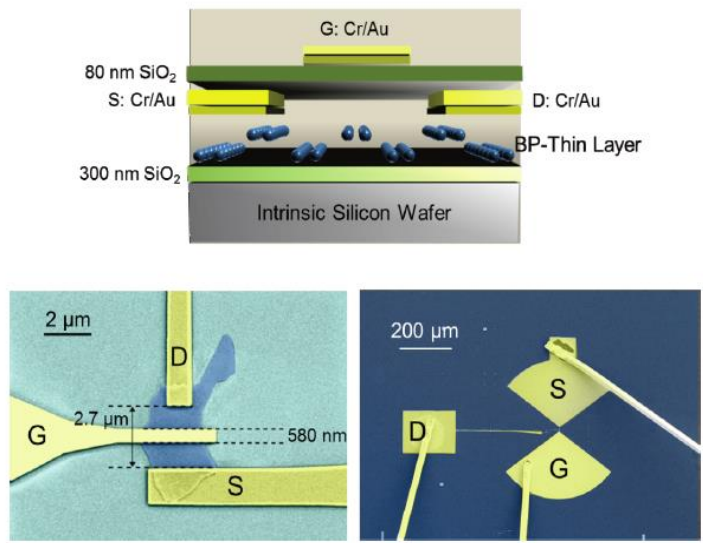

Figure 3

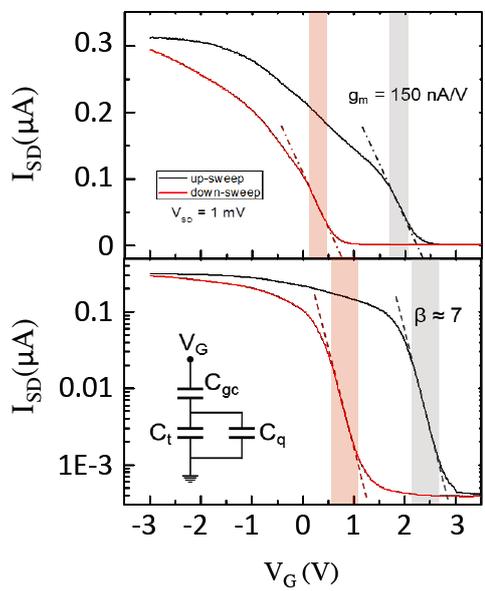

Figure 4

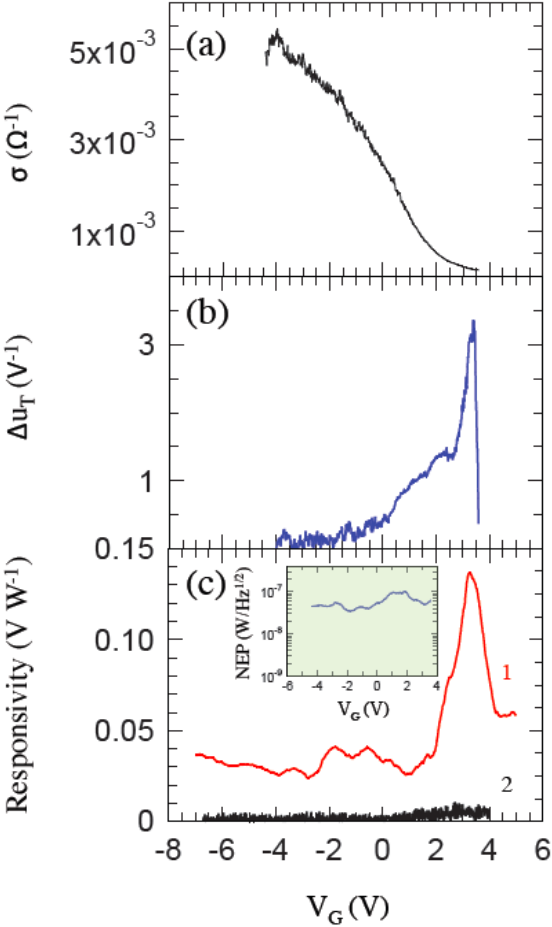